\documentclass[12pt]{article}
\usepackage{amssymb}
\usepackage{epsfig}
\usepackage{graphicx}
\usepackage{amsmath}
\usepackage{verbatim}

\csname @addtoreset\endcsname{equation}{section}
\begin{document}
\begin{titlepage}
\title{\bf\Large  Variant Supercurrents and Linearized Supergravity  \vspace{18pt}}

\author{\normalsize Sibo~Zheng$^{1}$~and~Jia-Hui~Huang$^{2}$ \vspace{12pt}\\
{\it\small  $^{1}$ Department of Physics, Chongqing University, Chongqing 400030, P.R. China}\\
{\it\small $^{2}$ Center of Mathematical Science, Zhejiang University, Hangzhou 310027, P.R. China}\\
}

\date{}
\maketitle \voffset -.3in \vskip 1.cm \centerline{\bf Abstract}
\vskip .3cm
In this paper the variant supercurrents based on consistency and completion in off-shell $\mathcal{N}=1$ supergravity are studied.
We formulate the embedding relations for supersymmetric current and energy tensor into supercurrent multiplet.
Corresponding linearized supergravity is obtained with appropriate choice of Wess-Zumino gauge in each gravity supermultiplet.
\vskip 5.cm \noindent July  2010
 \thispagestyle{empty}

\end{titlepage}
\newpage
\section{Introduction}
According to the structure of supersymmetry algebra,
the $R$ current $j^{5}_{\mu}$, supersymmetric current $j_{\mu}$ and energy tensor $T_{\mu\nu}$ corresponding to
the $R$ charge, supercharge and spacetime momentum respectively can be embedded into a supermultiplet.
This multiplet is known as supercurrent \cite{Ferrara}.
The superfield form of a supercurrent and the constraint it satisfies are found to be quite model dependent,
although some general considerations from symmetries can be taken into account \cite{OS1,OS2,Howe,CPS}.

There is a standard scheme for analyzing the structure of supercurrent and its corresponding
linearized supergravity for a given physical system.
The procedure is as follows:
\begin{enumerate}
  \item Begin with the physics systems studied, and the conservation conditions,
\begin{eqnarray}
\partial^{\mu}j_{\mu}=0,~~~~~~~\partial^{\mu}T_{\mu\nu}=0 \nonumber
\end{eqnarray}
one has to find the embedding relations for supersymmetric current and energy-momentum tensor into supercurrent.
During this stage the supercurrent multiplet and the constraint it satisfies are determined at meantime.
  \item Through the constraint that supercurrent satisfies we obtain the constraints on gauge transformation superfield
  $L$ of gravity supermultiplet, which tell us the analogy of Wess-Zumino gauge in gravity supermultiplet.
  \item Collect the embedding relations of gravity and gravitino into gravity supermultiplet together,
  the action of linearized supergravity can be directly read in components.
\end{enumerate}

In this paper we study the structures of three new variant supercurrent \cite{ Kuzenko} using the results obtained earlier in \cite{0108200, 0306288}.
The existence of these variant supercurrent is based on consistency and completion in $N=1$
off-shell linearized supergravity .
Other supercurrents
deduced via this viewpoint include the Ferrara-Zumino (FZ) multiplet,
new minimal multiplet \cite{West} and $S$ multiplets \cite{1002.2228}
all of which has completions of quantum field theories (see also \cite{0110131}).
The varaint supercurrents are defined as follows.
\begin{eqnarray}{\label{constraint1}}
Case~I:~~~~~\bar{D}^{\dot{\alpha}}J^{I}_{\alpha\dot{\alpha}}=i\eta_{\alpha},~~~~~
\bar{D}\eta=D^{\alpha}\eta_{\alpha}-\bar{D}^{\dot{\alpha}}\bar{\eta}_{\dot{\alpha}}=0
\end{eqnarray}
which is a minimal off-shell supergravity.
The second case is,
\begin{eqnarray}{\label{constraint2}}
Case~II:~~\bar{D}^{\dot{\alpha}}J^{II}_{\alpha\dot{\alpha}}=i\eta_{\alpha}+\hat{\chi}_{\alpha},
\end{eqnarray}
with $\bar{D}^{\dot{\alpha}}\eta_{\alpha}=D^{\alpha}\eta_{\alpha}-\bar{D}^{\dot{\alpha}}\bar{\eta}_{\dot{\alpha}}=0$ and
$\bar{D}^{\dot{\alpha}}\hat{\chi}_{\alpha}=D^{\alpha}\hat{\chi}_{\alpha}-\bar{D}^{\dot{\alpha}}\bar{\hat{\chi}}_{\dot{\alpha}}=0$.
The last case is,
\begin{eqnarray}{\label{constraint3}}
Case~III:~~~\bar{D}^{\dot{\alpha}}J^{III}_{\alpha\dot{\alpha}}=i\eta_{\alpha}+D_{\alpha}X
\end{eqnarray}
with $\bar{D}^{\dot{\alpha}}\eta_{\alpha}=D^{\alpha}\eta_{\alpha}-\bar{D}^{\dot{\alpha}}\bar{\eta}_{\dot{\alpha}}=0$
and $\bar{D}X=0$.

There are some common results in variant supercurrent.
Firstly, the $R$ current is not conserved,
which can be easily observed from the constraint of eq\eqref{constraint1} to eq\eqref{constraint3} .
Secondly, there exists some special constraints for energy tensor $T_{\mu\nu}$ as shown below.
These constraints exclude some simple physical systems we are familiar with.
Thus, they might serve as the necessary conditions for existence of variant supercurrent.

The paper is organized as follows. In section 2, we discuss the minimal case I.
Section 3 are denoted to study non-minimal cases II and III.
The solutions to the constraint \eqref{constraint1} to eq\eqref{constraint3} are obtained,
with comments on conditions that energy-tensor has to satisfy.
The actions of linearized supergravity are obtained after the analogy of Wess-Zumino gauge in each case are discussed.
In section 4, we conclude and discuss the difference between variant supercurrents and other supercurrents in the literature.

\section{Minimal Case I}
In this note we follow the conventions of Wess and Bagger \cite{Wess}.
The real vector superfield $J_{\mu}$ is defined in bi-spinor representation as
\begin{eqnarray}
J_{\alpha\dot{\alpha}}=\sigma^{\mu}_{\alpha\dot{\alpha}}J_{\mu},~~~~~~
and~~~~ J_{\mu}=-\frac{1}{2}\bar{\sigma}_{\mu}^{\dot{\alpha}\alpha}J_{\alpha\dot{\alpha}}.
\end{eqnarray}
The components are expressed as,
\begin{eqnarray}{\label{new}}
S&=&C^{S}+i\theta\chi^{S}-i\bar{\theta}\bar{\chi}^{S}+\frac{i}{2}\theta^{2}(M^{S}+iN^{S})-\frac{i}{2}\bar{\theta}^{2}(M^{S}-iN^{S})
-\theta\sigma^{m}\bar{\theta}\upsilon^{S}_{m}\nonumber\\
&+&i\theta^{2}\bar{\theta}\left(\bar{\lambda}^{S}+\frac{i}{2}\bar{\sigma}^{m}\partial_{m}\chi^{S}\right)
-i\bar{\theta}^{2}\theta\left(\lambda^{S}+\frac{i}{2}\sigma^{m}\partial_{m}\bar{\chi}^{S}\right)
+\frac{1}{2}\theta^{2}\bar{\theta}^{2}\left(D^{S}+\frac{1}{2}\Box~C^{S}\right)\nonumber\\
\end{eqnarray}
Note that the lowest component field $C^J$ in supercurrent superfield $J$ is the $R$ current $j^{5}_{\mu}$.

We deduce a new constraint on supercurrent from the constraint eq\eqref{constraint1},
\begin{eqnarray}{\label{self1}}
\bar{D}^{\dot{\beta}}\bar{D}^{\dot{\alpha}}J^{I}_{\alpha\dot{\alpha}}=0
\end{eqnarray}
The first equation in constraint eq.\eqref{constraint1} can be classified into its real and imaginary parts,
respectively\footnote{Similar methods are applied to the other two cases we will discuss in this note.}.
Explicit expressions for these components can be found in \cite{0911.0677}.

Solving eq\eqref{self1} and eq\eqref{constraint1} we obtain,
\begin{eqnarray}{\label{J1}}
J_{\mu}^{I}&=&C_{\mu}+\theta\left(j_{\mu}+\frac{1}{3}\sigma_{\mu}\bar{\sigma}^{\nu}j_{\nu}\right)
+\bar{\theta}\left(\bar{j}_{\mu}+\frac{1}{3}\bar{\sigma}_{\mu}\sigma^{\nu}\bar{j}_{\nu}\right)\nonumber\\
&+&(\theta\sigma^{\nu}\bar{\theta})\left(aT_{\nu\mu}+bT\eta_{\nu\mu}+\frac{1}{4}
\epsilon_{\nu\mu\rho\lambda}\left(\partial^{\rho}C^{\lambda}-\partial^{\lambda}C^{\rho}\right)-\frac{1}{2}\Phi_{\nu\mu}\right)\\
&+&\theta^{2}\bar{\theta}\left(\frac{i}{3}\bar{\sigma}^{\nu}\partial_{\mu}j_{\nu}\right)
+\bar{\theta}^{2}\theta\left(\frac{i}{3}\sigma^{\nu}\partial_{\mu}\bar{j}_{\nu}\right)
+\bar{\theta}^{2}\theta^{2}\left(-\frac{1}{2}\Box~C_{\mu}-\frac{1}{2}\partial_{\mu}\partial^{\nu}C_{\nu}\right)\nonumber
\end{eqnarray}
and
\begin{eqnarray}
\eta_{\alpha}&=&-i\Lambda_{\alpha}(y)+\left(\delta^{\beta}_{\alpha}\Delta-2i\bar{\sigma}^{\mu}\sigma^{\nu}\Phi_{\mu\nu}(y)\right)\theta_{\beta}
+\theta^{2}(\sigma^{\mu}\partial_{\mu}\bar{\Lambda}(y))_{\alpha}
\end{eqnarray}
where the coefficient $a,~b$ is introduced to define $\hat{T}_{\mu}\mid_{s}=aT_{\mu\nu}+b\eta_{\mu\nu}T$.
In this case, constant $a$, $b$ are given by,
\begin{eqnarray}{\label{b}}
a=-4b,~~~~2b\partial_{\nu}T=-\partial^{\mu}\Phi_{\mu\nu},~~~~~~\Box~T=0
\end{eqnarray}
The lower indices $s, a$ in $\hat{T}_{\mu\nu}\mid_{}$ refer to the
symmetric and anti-symmetric part respectively. $\Phi^{\rho\sigma}$ and $\Delta$
is the tensor field and $D$-term in $\eta$ superfield respectively. The degrees of freedom of
$\hat{T}_{\mu\nu}\mid_{a}$ can be considered as totally provided by
$\Phi^{\rho\sigma}$ .  Physical systems with energy tensor
$T_{\mu\nu}$ that satisfies these special constraints are
extraordinary. The non-existence of these conditions might serve as
a proof that the first kind of constrained supercurrent is not
physical. This question will be investigated further.

The degrees of freedom in this case are described by
$(C_{\mu}, \chi_{\mu},\hat{T}_{\mu\nu}\mid_{s}, \hat{T}_{\mu\nu}\mid_{a})$,
which imply supersymmetric theories correspond to 12/12 off-shell supergravity.
Gauging the supercurrent $J^I$ in supergravity via coupling
\begin{eqnarray}{\label{action1}}
\int d^{4}x \int d^{4}\theta~ J^{I}_{\alpha\dot{\alpha}}H^{\alpha\dot{\alpha}}
\end{eqnarray}
Gauge invariance of action eq\eqref{action1} under transformation $H_{\mu}\rightarrow~H_{\mu}+\triangle_{\mu}$, or
equivalently via its bi-spinor expression
\begin{eqnarray}{\label{a}}
H_{\alpha\dot{\alpha}}\rightarrow~H_{\alpha\dot{\alpha}}+D_{\alpha}\bar{L}_{\dot{\alpha}}
-\bar{D}_{\dot{\alpha}}L_{\alpha}
\end{eqnarray}
leads to
\begin{eqnarray}{\label{gauge1}}
\bar{D}_{\dot{\alpha}}D^{2}\bar{L}^{\dot{\alpha}}+D_{\alpha}\bar{D}^{2}L^{\alpha}=0
\end{eqnarray}
Here superfield $L$ is defined as,
\begin{eqnarray}{\label{L}}
\triangle_{\mu}=-\frac{1}{2}\bar{\sigma}_{\mu}^{\dot{\alpha}\alpha}\left(D_{\alpha}\bar{L}_{\dot{\alpha}}-\bar{D}_{\dot{\alpha}}L_{\alpha}\right)
\end{eqnarray}
$\triangle_{\mu}$ is a general real superfield.
Eq\eqref{L} suggests that the relations of embedding graviton and gravitino into supergravity multiplet $H_{u}$
follow those of \cite{Weinberg,1002.2228}
\footnote{Following the conventions we take, one can see
 these embedding relations are independent of constraints on $L_{\alpha}$.
They are valid throughout this note.}.
$H_{\mu}\mid_{\theta\bar{\theta}}$ is divided into the symmetric
part $\upsilon^{H}_{\mu\nu}$ and anti-symmetric part $B_{\mu\nu}$.
The gauge transformations are as follows,
\begin{eqnarray}
\delta~h_{\mu\nu}=\partial_{\mu}\xi_{\nu}+\partial_{\nu}\xi_{\mu},~~~~~
\delta~\Psi_{\mu\alpha}=\partial_{\mu}\omega_{\alpha}\nonumber
\end{eqnarray}

The constraint Eq\eqref{gauge1} impose some equations in components in $L$,
which implies a set of constraint equations in components of $\triangle_{\mu}$ via eq\eqref{L}.
These constraints determine the analog of the Wess-Zumino gauge for supermultiplet $H_{\mu}$.
Define
\begin{eqnarray}
L_{\alpha}=iD_{\alpha}V
\end{eqnarray}
Eq\eqref{gauge1} leads to the identification of $V$ as Wess-Zumino gauged vector superfield.
The constraints on components in $\triangle_{\mu}$ are,
\begin{eqnarray}
L_{\alpha}\mid=L_{\alpha}\mid_{\theta}&=&L_{\alpha}\mid_{\theta^{2}}=L_{\alpha}\mid_{\theta^{2}\bar{\theta}}=0
\end{eqnarray}
and
\begin{eqnarray}
\partial^{m}(L_{\alpha}\mid_{\theta\sigma^{m}\bar{\theta}})&=&-2(L_{\alpha}\mid_{\theta^{2}\bar{\theta}^{2}})
\end{eqnarray}
which imply that $B_{\mu\nu}$ field in gravity supermultiplet can not be set to zero.

One can see that the analogy of Wess-Zumino gauge is as follows,
\begin{eqnarray}{\label{WZ}}
H_{\mu}\mid=H_{\mu}\mid_{\theta}=H_{\mu}\mid_{\bar{\theta}}=H_{\mu}\mid_{\theta^{2}}=H_{\mu}\mid_{\bar{\theta}^{2}}=0
\end{eqnarray}
The residual degrees of freedom in gravity supermultiplet are represented by
$(h_{\mu\nu}$,$B_{\mu\nu}$,$\Psi_{\mu\alpha}$ and $D^{H}_{\mu})$,
which describe 12/12 minimal supergravity.
They match with that of supercurrent.

Following notation eq.\eqref{new}, we obtain the action in components,
\begin{eqnarray}{\label{s1}}
S=-\upsilon_{\mu\nu}^{H}\hat{T}^{\mu\nu}\mid_{s}-B_{\mu\nu}\hat{T}^{\mu\nu}\mid_{a}
+\frac{1}{2}j^{5\mu}D^{H}_{\mu}+\left(\frac{i}{2}\chi_{\mu}^{(J)}\lambda^{(H)\mu}+c.c\right)
\end{eqnarray}
The kinetic term of graviton can be constructed via appropriate derivative operator \cite{Kuzenko}.
Starting with the constraint on gauge transformation superfield, the results in \cite{Kuzenko} can be reproduced.
Similar results can be found in non-minimal cases.

\section{Reducible Cases}
Now we discuss the non-minimal case II and case III.
Their supercurrent multiplets both include $16+16$ degrees of freedom (supermultiplets with $16+16$ degrees of freedom
are also discussed in \cite{LLO,Siegel}),
which are  manifested by their corresponding gravity supermultiplets.
In comparison with the minimal case I,
the gauge transformation superfield $L_{\alpha}$ is more constrained,
which is the origin of more degrees of freedom in gravity supermultiplets.

\subsection{Reducible Cases II}
The constraint eq\eqref{constraint2} implies that,
\begin{eqnarray}{\label{self2}}
\bar{D}^{\dot{\beta}}\bar{D}^{\dot{\alpha}}J^{II}_{\alpha\dot{\alpha}}=0
\end{eqnarray}
Solving eq\eqref{self2} and eq\eqref{constraint2} gives,
\begin{eqnarray}{\label{J2}}
J_{\mu}^{II}&=&C_{\mu}+\theta\left(j_{\mu}+\frac{1}{3}\sigma_{\mu}\bar{\sigma}^{\nu}j_{\nu}+\frac{1}{3}\sigma_{\mu}\bar{\psi}\right)
+\bar{\theta}\left(\bar{j}_{\mu}+\frac{1}{3}\bar{\sigma}_{\mu}\sigma^{\nu}\bar{j}_{\nu}-\frac{1}{3}\bar{\sigma}_{\mu}\psi\right)\nonumber\\
&+&(\theta\sigma^{\nu}\bar{\theta})\left(aT_{\nu\mu}-\frac{b}{a+4b}Z\eta_{\nu\mu}+\frac{1}{4}
\epsilon_{\nu\mu\rho\lambda}\left(\partial^{\rho}C^{\lambda}-\partial^{\lambda}C^{\rho}+\Sigma^{\rho\lambda}\right)-\frac{1}{2}\Phi_{\nu\mu}\right)\\
&+&\theta^{2}\bar{\theta}\left(-\frac{2i}{3}\partial_{\mu}\bar{\psi}+\frac{i}{3}\bar{\sigma}^{\nu}\partial_{\mu}j_{\nu}\right)
+\bar{\theta}^{2}\theta\left(\frac{2i}{3}\partial_{\mu}\psi+\frac{i}{3}\sigma^{\nu}\partial_{\mu}\bar{j}_{\nu}\right)\nonumber\\
&+&\bar{\theta}^{2}\theta^{2}\left(\frac{1}{2}\partial_{\mu}Z-\frac{1}{2}\Box~C_{\mu}-\frac{1}{2}\partial_{\mu}\partial^{\nu}C_{\nu}+\frac{3}{2}\partial^{\nu}\Sigma_{\mu\nu}\right)\nonumber
\end{eqnarray}
and
\begin{eqnarray}
\eta_{\alpha}&=&-i\Lambda_{\alpha}(y)+\left(\delta^{\beta}_{\alpha}\Delta-2i\bar{\sigma}^{\mu}\sigma^{\nu}\Phi_{\mu\nu}(y)\right)\theta_{\beta}
+\theta^{2}(\sigma^{\mu}\partial_{\mu}\bar{\Lambda}(y))_{\alpha}\nonumber\\
\hat{\chi}_{\alpha}&=&-i\psi_{\alpha}(y)+\left(\delta^{\beta}_{\alpha}Z-2i\bar{\sigma}^{\mu}\sigma^{\nu}\Sigma_{\mu\nu}(y)\right)\theta_{\beta}
+\theta^{2}(\sigma^{\mu}\partial_{\mu}\bar{\psi}(y))_{\alpha}
\end{eqnarray}
The coefficient $a,~b$ satisfy
\begin{eqnarray}{\label{b2}}
\left(a+4b\right)T=-Z,~~~~2b\partial_{\nu}T=-\partial^{\mu}\Phi_{\mu\nu}
\end{eqnarray}
As emphasized above, the existence of $a,~b$ is necessary for physical systems described by the case II.
The multiplet $J_{\mu}^{II}$ contain $12+12$ degrees of freedom, a Weyl spinor $\psi$, a closed two-form $\Sigma_{\mu\nu}$,
and a real scalar $Z$. Thus it describes $16+16$ supermultiplet.

Gauging the supercurrent $J^{II}$ in supergravity via coupling
\begin{eqnarray}{\label{action2}}
\int d^{4}x \int d^{4}\theta~ J^{II}_{\alpha\dot{\alpha}}H^{\alpha\dot{\alpha}}
\end{eqnarray}
Gauge invariance of the action under transformation eq.\eqref{a} leads to
\begin{eqnarray}{\label{gauge2}}
\bar{D}^{\dot{\alpha}}D^{2}\bar{L}_{\dot{\alpha}}=D^{\alpha}\bar{D}^{2}L_{\alpha}=0
\end{eqnarray}
The embedding relations of graviton and gravitino into $H_{\mu}$ superfield is the same as in case I.
Note that the equation of motion of a field strength chiral superfield without FI term
is exactly the same with eq\eqref{gauge2}.
The analogy of Wess-Zumino gauge is given by,
\begin{eqnarray}{\label{WZ}}
H_{\mu}\mid=H_{\mu}\mid_{\theta}=H_{\mu}\mid_{\bar{\theta}}=H_{\mu}\mid_{\theta^{2}}=H_{\mu}\mid_{\bar{\theta}^{2}}=0
\end{eqnarray}
The residual degrees of freedom in gravity supermultiplet are represented by
$(h_{\mu\nu}$,$B_{\mu\nu}$,$\Psi_{\mu\alpha}$ and $D^{H}_{\mu})$, which describe 16/16 linearized supergravity.
They match with that of supercurrent.
Corresponding action is in components with notation eq.\eqref{new},
\begin{eqnarray}{\label{s2}}
S=-\upsilon_{\mu\nu}^{H}\hat{T}^{\mu\nu}\mid_{s}-B_{\mu\nu}\hat{T}^{\mu\nu}\mid_{a}
+\frac{1}{2}j^{5\mu}D^{H}_{\mu}+\left(\frac{i}{2}\chi_{\mu}^{(J)}\lambda^{(H)\mu}+c.c\right)
\end{eqnarray}

\subsection{Reducible Cases III}
Finally we address the third possible constraint satisfied by supercurrent.
Solving equation eq.\eqref{constraint3} we obtain $J_{\mu}^{III}$,
\begin{eqnarray}{\label{J3}}
J_{\mu}^{III}&=&C_{\mu}+\theta\left(j_{\mu}+\frac{1}{3}\sigma_{\mu}\bar{\sigma}^{\nu}j_{\nu}\right)
+\bar{\theta}\left(\bar{j}_{\mu}+\frac{1}{3}\bar{\sigma}_{\mu}\sigma^{\nu}\bar{j}_{\nu}\right)
-i\theta^{2}\partial_{\mu}\phi+i\bar{\theta}^{2}\partial_{\mu}\phi^{*}\nonumber\\
&+&(\theta\sigma^{\nu}\bar{\theta})\left(aT_{\nu\mu}-2Re(F)\eta_{\nu\mu}+\frac{1}{2}
\epsilon_{\nu\mu\rho\lambda}\partial^{\rho}C^{\lambda}-\frac{1}{2}\Phi_{\nu\mu}\right)\\
&+&\theta^{2}\bar{\theta}\left(\frac{i}{3}\bar{\sigma}^{\rho}\partial_{\mu}j_{\rho}-\sqrt{2}\partial_{\mu}\bar{\psi}\right)
+\bar{\theta}^{2}\theta\left(\frac{i}{3}\sigma^{\rho}\partial_{\mu}\bar{j}_{\rho}-\sqrt{2}\partial_{\mu}\psi\right)\nonumber\\
&+&\bar{\theta}^{2}\theta^{2}\left(-2\partial_{\mu}(Im(F))+\frac{1}{2}\Box~C_{\mu}-\frac{3}{2}\partial_{\mu}\partial^{\rho}C_{\rho}\right)\nonumber
\end{eqnarray}
and
\begin{eqnarray}
X&=&\phi(y)+\sqrt{2}\theta\psi(y)+\theta^{2}F\nonumber\\
\eta_{\alpha}&=&-i\Lambda_{\alpha}(y)+\left(\delta^{\beta}_{\alpha}\Delta-2i\bar{\sigma}^{\mu}\sigma^{\nu}\Phi_{\mu\nu}(y)\right)\theta_{\beta}
+\theta^{2}(\sigma^{\mu}\partial_{\mu}\bar{\Lambda}(y))_{\alpha}
\end{eqnarray}
The components fields in $\eta_{\alpha}$ satisfy extra constraints,
\begin{eqnarray}
\Delta&=&-\partial^{\mu}C_{\mu}-2Im(F), \nonumber\\
\Lambda_{\alpha}&=&\frac{i}{3}(\sigma^{\mu}\bar{j}_{\mu})_{\alpha}-\sqrt{2}\psi_{\alpha}
\end{eqnarray}
Here the coefficient $a$ is given by $aT=6Re(F)$, with $F=Re(F)+iIm(F)$.
The multiplet $J_{\mu}^{III}$ contains $12+12$ degrees of freedom, a Weyl spinor $\psi$, a complex scalar $\phi$,
and a complex scalar $F$ (or equivalently $Re(F)$ and $\Delta$), which imply that it is actually $16+16$ supermultiplet.
Compared with the $S$-multiplet that is introduced to solve problem of FI term in supergravity \cite{1002.2228},
the scalar $Re(F)$ is now replaced by $F$.
The embedding realtions are also very different.

Gauging the supercurrent $J^{III}$ in supergravity via coupling
\begin{eqnarray}{\label{action3}}
\int d^{4}x \int d^{4}\theta~ J^{III}_{\alpha\dot{\alpha}}H^{\alpha\dot{\alpha}}
\end{eqnarray}
Gauge invariance of the action under transformation  eq.\eqref{a} leads to
\begin{eqnarray}{\label{gauge3}}
\bar{D}^{2}D^{\alpha}L_{\alpha}=0,~~~
\bar{D}_{\dot{\alpha}}D^{2}\bar{L}^{\dot{\alpha}}=D_{\alpha}\bar{D}^{2}L^{\alpha}
\end{eqnarray}
As more constraints are imposed, less component fields in gravity supermultiplet can be
set to zero. The constraint eq\eqref{gauge3} suggests that the analog of Wess-Zumino guage is
\begin{eqnarray}{\label{WZ3}}
H_{\mu}\mid=H_{\mu}\mid_{\theta}=H_{\mu}\mid_{\bar{\theta}}=0
\end{eqnarray}
The action can be read in components with notation eq.\eqref{new},
\begin{eqnarray}{\label{s3}}
S=-\upsilon_{\mu\nu}^{H}\hat{T}^{\mu\nu}\mid_{s}-B_{\mu\nu}\hat{T}^{\mu\nu}\mid_{a}
+\frac{1}{2}j^{5\mu}D^{H}_{\mu}+\left[\frac{i}{2}\chi_{\mu}^{(J)}\lambda^{(H)\mu}
+\frac{1}{4}\left(M^{J}+iN^{J}\right)\left(M^{H}-iN^{H}\right)+c.c\right]\nonumber\\
\end{eqnarray}

\section{Conclusions}
In this note we study a set of variant supercurrents that arise from consistency and completion
 in $\mathcal{N}=1$ off-shell supergravity.
We use the component languages of superfield to obtain the embedding relations of supersymmetric current
and  energy-momentum tensor into formalism of linear supergravity.
The analogy of Wess-Zumino gauge in each case is analyzed in details.

Instead of the superfield formalisms used to describe variant supercurrents,
we find more physical results are uncovered in the component expressions.
First, the consistent conditions for energy-momentum tensor of supersymmetric theories that can be described by
variant supercurrent multiplets are determined explicitly.
Second, the component results help identifying corresponding linearized supergravity.

Although supercurrents that include $S$-multiplet \cite{1002.2228}, FZ-multiplet and minimal mutiplet
have rich constructions of quantum field theories and
important applications\footnote{Recently, it is found in \cite{1002.2228} that $S$-multiplet is useful to embed
the Fayet-Iliopoulos term for abelian gauge supermultiplet into supergravity. },
the consistent conditions eq.\eqref{b} and eq.\eqref{b2} for variant supercurrents
studied in this paper imply $\Box~T=0$.
It can be verified that the variant supercurrents are not viable for simple supersymmetric field theories
including pure supersymmetric Yang-Mills theories and SQCD-like theories as a result of $\Box~T\neq 0$.
The main reason for this difference between variant supercurrent and $S$-multiplet is that
the $i$ factor in front of  the linear superfield in eq.\eqref{constraint1} to eq.\eqref{constraint3}
leads to the real and imaginary part of ${\theta^{2}\bar{\theta}}$ component
in these constraint equations exchanged.
In the case of $S$-multiplet,
the energy momentum tensor depends on $D$-term of $\eta_{\alpha}$ superfield
\footnote{As shown in \cite{1002.2228},  $\Box~T$ is proportional to
$\Box~Re~F_{X}$ and $\Box~\Delta$.
Although the  component expressions of $F$ and $D$ terms are quite involved,  it is expected that both $\Box~Re~F_{X}\neq 0$ and $\Box~\Delta\neq 0$ in terms of equations of motion of relevant fields.
In other words,  $\Box~T=0$ is forbidden in case of $S$ multiplet. },
and anti-symmetric tensor field $\Phi_{\mu\nu}$ is related to
the anti-symmetric part of $\hat{T}_{\mu\nu}$ .
In the case of variant supercurrents, however,
exchanging the real and imaginary parts of component ${\theta^{2}\bar{\theta}}$
in eq.\eqref{constraint1} to eq.\eqref{constraint3}
leads to that the dependence on anti-symmetric tensor field $\Phi_{\mu\nu}$
is transferred to the derivative of energy-momentum tensor trace $\partial_{\mu}T$ ,  which
is the origin of the severe constraint $\Box~T=0$ for variant supercurrents.

\section*{Acknowledgement}
This work is supported in part by
the Fundamental Research Funds for the Central
Universities with project number CDJRC10300002.

\end{document}